\documentclass[twoside,twocolumn,9pt]{article}
\usepackage{extsizes}
\usepackage[super,sort&compress,comma]{natbib} 
\usepackage[version=3]{mhchem}
\usepackage[left=1.5cm, right=1.5cm, top=1.785cm, bottom=2.0cm]{geometry}
\usepackage{balance}
\usepackage{times,mathptmx}
\usepackage{sectsty}
\usepackage{graphicx} 
\usepackage{lastpage}
\usepackage[format=plain,justification=justified,singlelinecheck=false,font={stretch=1.125,small,sf},labelfont=bf,labelsep=space]{caption}
\usepackage{float}
\usepackage{fancyhdr}
\usepackage{fnpos}
\usepackage[english]{babel}
\addto{\captionsenglish}{%
  
}
\usepackage{array}
\usepackage{droidsans}
\usepackage{charter}
\usepackage[T1]{fontenc}
\usepackage[usenames,dvipsnames]{xcolor}
\usepackage{setspace}
\usepackage[compact]{titlesec}
\usepackage{hyperref}
\usepackage{chemscheme}
\usepackage{gensymb}

\definecolor{cream}{RGB}{222,217,201}

\begin{document}

\pagestyle{fancy}
\thispagestyle{plain}
\fancypagestyle{plain}{

\renewcommand{\headrulewidth}{0pt}
}

\makeFNbottom
\makeatletter
\renewcommand\LARGE{\@setfontsize\LARGE{15pt}{17}}
\renewcommand\Large{\@setfontsize\Large{12pt}{14}}
\renewcommand\large{\@setfontsize\large{10pt}{12}}
\renewcommand\footnotesize{\@setfontsize\footnotesize{7pt}{10}}
\makeatother

\renewcommand{\thefootnote}{\fnsymbol{footnote}}
\renewcommand\footnoterule{\vspace*{1pt}%
\color{cream}\hrule width 3.5in height 0.4pt \color{black}\vspace*{5pt}} 
\setcounter{secnumdepth}{5}

\makeatletter 
\renewcommand\@biblabel[1]{#1}            
\renewcommand\@makefntext[1]%
{\noindent\makebox[0pt][r]{\@thefnmark\,}#1}
\makeatother 
\renewcommand{\figurename}{\small{Fig.}~}
\sectionfont{\sffamily\Large}
\subsectionfont{\normalsize}
\subsubsectionfont{\bf}
\setstretch{1.125} 
\setlength{\skip\footins}{0.8cm}
\setlength{\footnotesep}{0.25cm}
\setlength{\jot}{10pt}
\titlespacing*{\section}{0pt}{4pt}{4pt}
\titlespacing*{\subsection}{0pt}{15pt}{1pt}

\makeatletter 
\newlength{\figrulesep} 
\setlength{\figrulesep}{0.5\textfloatsep} 

\newcommand{\topfigrule}{\vspace*{-1pt}%
\noindent{\color{cream}\rule[-\figrulesep]{\columnwidth}{1.5pt}} }

\newcommand{\botfigrule}{\vspace*{-2pt}%
\noindent{\color{cream}\rule[\figrulesep]{\columnwidth}{1.5pt}} }

\newcommand{\dblfigrule}{\vspace*{-1pt}%
\noindent{\color{cream}\rule[-\figrulesep]{\textwidth}{1.5pt}} }

\makeatother

  \begin{@twocolumnfalse}
\vspace{3cm}
\sffamily

\noindent\LARGE{\textbf{A study of the ozonolysis of isoprene in a cryogenic buffer gas cell by high resolution microwave spectroscopy}} \\


\noindent\large{Jessica P. Porterfield,\textit{$^{a*}$} Sandra Eibenberger,\textit{$^{a\ddag}$} Dave Patterson,\textit{$^{b}$} and Michael C. McCarthy\textit{$^{a}$}} \\

\noindent\normalsize{We have developed a method to quantify reaction product ratios using high resolution microwave spectroscopy in a cryogenic buffer gas cell. We demonstrate the power of this method with the study of the ozonolysis of isoprene, \ce{CH2=C(CH3)-CH=CH2}, the most abundant, non-methane hydrocarbon emitted into the atmosphere by vegetation. 
Isoprene is an asymmetric diene, and reacts with \ce{O3} at the 1,2 position to produce methyl vinyl ketone (MVK), formaldehyde, and a pair of carbonyl oxides: [\ce{CH3CO-CH=CH2 + CH2=OO}] + [\ce{CH2=O + CH3COO-CH=CH2}]. 
Alternatively, \ce{O3} could attack at the 3,4 position to produce methacrolein (MACR), formaldehyde, and two carbonyl oxides [\ce{CH2=C(CH3)-CHO + CH2=OO}] + [\ce{CH2=O + CH2=C(CH3)-CHOO}]. 
Purified \ce{O3} and isoprene were mixed for approximately 10 seconds under dilute (1.5-4\% in argon) continuous flow conditions in an alumina tube held at 298 K and 5 Torr.
Products exiting the tube were rapidly slowed and cooled within the buffer gas cell by collisions with cryogenic (4-7 K) He.
High resolution chirped pulse microwave detection between 12 and 26 GHz was used to achieve highly sensitive (ppb scale), isomer-specific product quantification. 
We observed a ratio of MACR to MVK of 2.1 $\pm$ 0.4 under 1:1 ozone to isoprene conditions and 2.1 $\pm$ 0.2 under 2:1 ozone to isoprene conditions, a finding which is consistent with previous experimental results. 
Additionally, we discuss relative quantities of formic acid (HCOOH), an isomer of \ce{CH2=OO}, and formaldehyde (\ce{CH2=O}) under varying experimental conditions, and characterize the spectroscopic parameters of the singly-substituted $^{13}$C $trans$-isoprene and $^{13}$C $anti$-periplanar-methacrolein species. 
This work has the potential to be extended towards a complete branching ratio analysis, as well towards the ability to isolate, identify, and quantify new reactive intermediates in the ozonolysis of alkenes.} \\


 \end{@twocolumnfalse} \vspace{0.6cm}

\renewcommand*\rmdefault{bch}\normalfont\upshape
\rmfamily
\section*{}
\vspace{-1cm}

\footnotetext{\textit{$^{a}$~Harvard-Smithsonian Center for Astrophysics, Harvard University, Cambridge, MA 02138, USA. Tel: +617-495-9848; $^*$E-mail: jessica.porterfield@cfa.harvard.edu}}
\footnotetext{\textit{$^{b}$~Department of Physics, University of California Santa Barbara, Santa Barbara, CA 93106, USA }}

\footnotetext{\ddag~Present address: Fritz Haber Institute of the Max Planck Society, Berlin, Germany.}

\section{Introduction}
Isoprene is the most abundant non-methane hydrocarbon emitted into the atmosphere.
Approximately 500 - 750 Tg of isoprene are emitted by vegetation each year, with roughly 10\% removed globally by reaction with ozone \cite{guenther2006estimates, nguyen2016atmospheric}. 
The ozonolysis of isoprene converts a relatively inert, non-polar species into stable, polar products as well as three highly unstable Criegee intermediates, see Scheme 1. 
These intermediates tend to rapidly fragment, thus making them an important night time source of OH radicals, a key oxidant in the atmosphere.
Additionally it is believed that the stable products, methacrolein (MACR), methyl vinyl ketone (MVK), formaldehyde, and formic acid, aggregate and lead to secondary organic aerosol formation \cite{ervens2008secondary,atkinson2003gas, warneke2001isoprene}. 
Secondary aerosols give rise to clouds and chemical smog, and ultimately have an affect on climate and the quality of the air that we breathe \cite{lee2006gas}. 

In this article we describe a new method for product ratio analysis, and use this method to quantify the stable products of isoprene ozonolysis with great simplification relative to past experiments. Previous reports of the [MACR]/[MVK] ratio at 298 K span a large range (2.3 - 2.9), ostensibly because they entail long reaction times (hours) to obtain detectable quantities of products, and consequently require cumbersome calculations using variable kinetic data to compensate for secondary chemistry \cite{aschmann1994formation,grosjean1993atmospheric,rickard1999oh,nguyen2016atmospheric,ren2017investigation,kamens1982ozone}.
Here we quantify products in the nascent steps of the ozonolysis of isoprene within the first 10 seconds of the reaction. This dramatic reduction in time scale has been made possible by development of an instrument which possesses unique capabilities: high sensitivity (ppb scale), high resolution (50 kHz bandwidth), rapid data acquisition (50 kHz), continuous flow, and long observation times (15-20 ms).
Microwave diagnostics are being used in a new and powerful way, as demonstrated here as well as in product quantification of photolysis events \cite{zaleski2017time}, and as such shows promise of being a versatile analytical tool.

Products of the ozonolysis reaction were collisionally cooled for chirp-pulsed Fourier-transform microwave (CP-FTMW) detection between 12 and 26 GHz, a region where many rotational lines of stable oxidation products can be observed and relative abundances quantified. 
As a part of this work, lines of rare isotopic species are routinely observed for abundantly produced species, yielding line lists and spectroscopic constants when possible (e.g. singly substituted $^{13}$C $trans$-isoprene and $^{13}$C $anti$-periplanar ($ap$)-MACR). 
Finally, we discuss the prospects that the present work can be extended to derive a full branching ratio analysis with simple $^{13}$C isotopic substitution, as well as to detect reactive species such as the Criegee intermediates that have been proposed.

It is widely accepted that the primary pathway in the ozonolysis of unsaturated alkenes is the Criegee mechanism \cite{nguyen2016atmospheric,johnson2008gas, kuwata2005quantum, gutbrod1997kinetic, zhang2002mechanism,wadt1975electronic,womack2015observation}, see Scheme \ref{isoprene}.
This initially proceeds via formation of a primary ozonide (POZ) by addition of \ce{O3} to a C=C double bond, generating either POZ$_1$ (4-(2-propenyl)-1,2,3-trioxolane) or POZ$_2$ (4-methyl-4-vinyl-1,2,3-trioxolane).
Since formation of the primary ozonides is a highly exothermic process\cite{harding1978mechanisms} ($\Delta$H$_\textrm{rxn}$ approximately -200 kJ/mol), formation is proceeded by rapid decomposition to both a stable oxidized product and a highly reactive carbonyl oxide called a Criegee intermediate \cite{criegee1975mechanism}. There are two proposed Criegee intermediates in this reaction besides the simplest carbonyl oxide (\ce{CH2=OO}) which have never been observed: MVK-oxide (\ce{CH3C(OO)-CH=CH2}) and MACR-oxide (\ce{CH2=C(CH3)-CHOO}).
Because the exothermicity of the reaction is so great, these intermediates are energized, and they can rapidly rearrange and fragment to produce OH radicals among other species. As a consequence, their direct detection has proven to be very challenging. 
\begin{scheme}[h!]
\centering
\includegraphics[width=0.45\textwidth]{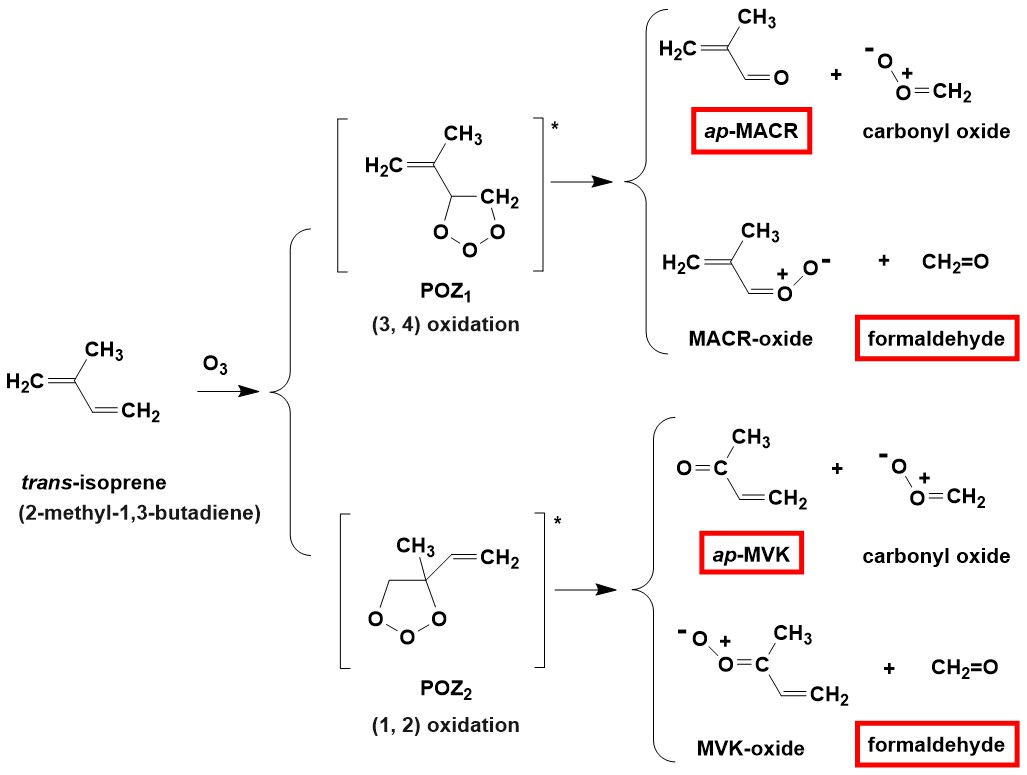}
\caption{The mechanism for ozonolysis of $trans$-isoprene. All three of the initial stable products: methyl vinyl ketone (\ce{CH3COCH=CH2}), methacrolein (\ce{CHOC(CH3)=CH2}), and formaldehyde (\ce{CH2=O}), have been characterized in this work.}
\label{isoprene}
\end{scheme}
\begin{scheme}[h!]
\centering
\caption{Conformers of species involved in the ozonolysis of isoprene, each of which has a unique rotational spectrum. The more stable conformers are listed on the left.}
\includegraphics[width=0.475\textwidth]{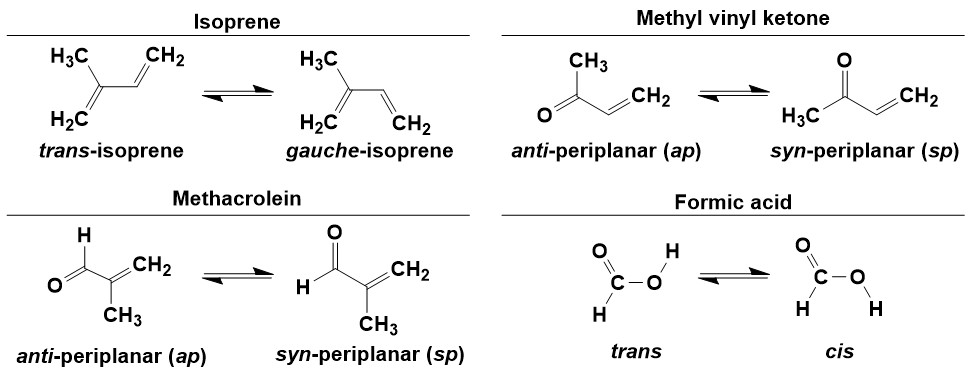}
\label{conformers}
\end{scheme}

If \ce{O3} addition occurs at the unsubstituted double bond of isoprene, the resulting POZ$_1$ could cleave in two different ways. 
The first path would lead to the formation of $ap$-MACR and the simplest Criegee intermediate, carbonyl oxide (\ce{CH2=OO}). 
Carbonyl oxide can be written as a zwitterion, \cite{su2013infrared,nakajima2013communication}, and although MACR-oxide and MVK-oxide are yet to be observed, they are also expected to be zwitterions \cite{kuwata2005quantum, kuwata2008quantum}.
The alternative pathway for POZ$_1$ involves elimination of formaldehyde, which leaves MACR oxide.
If instead ozone adds to the substituted double bond, it yields POZ$_2$.
Elimination of $ap$-MVK results in the formation of carbonyl oxide.
Alternatively loss of formaldehyde leads to the third Criegee intermediate accessible by this mechanism, MVK oxide. 
It should be noted that $trans$-isoprene as opposed to $gauche$-isoprene is considered here as a result of the room temperature distribution, which heavily favors the $trans$ conformer (roughly 98\%, see below).
The preponderance of the $trans$ conformer under our experimental conditions leads to rotamer specific oxidation products, $ap$-MVK and $ap$-MACR.
If one were to consider the ozonolysis of $gauche$-isoprene, we would expect the opposite rotamers (syn-periplanar, $sp$) to result based upon the mechanism presented in Scheme 1, i.e. see Scheme \ref{conformers}. 

The fate of the carbonyl oxides is complex.
Once the simplest carbonyl oxide (\ce{CH2=OO}) is formed, it can either be collisionally stabilized, or it can undergo unimolecular rearrangement \cite{womack2015observation,neeb1997formation}.
Rearrangement could involve a hydrogen transfer to hydroperoxymethylene (:CHOOH), followed by fragmentation to OH hydroxyl radical and HCO (ultimately H atoms and CO). 
It could instead rearrange via the hot ester mechanism, see Scheme \ref{formicacid}.
This involves ring closure to dioxirane (cyclic \ce{CH2OO}), ring opening to dioxymethylene (\ce{OCH2O}), and H atom transfer to form formic acid HCOOH.
Some fraction of the energized formic acid will ultimately dissociate \cite{goddard1992decarboxylation, wadt1975electronic, suenram1978dioxirane, neeb1997formation}.
An important consideration of \ce{CH2=OO} formation in the ozonolysis of isoprene is that the carbonyl products (MVK and MACR) produced in tandem are relatively large molecules, and are hence capable of carrying a significant fraction of the excess internal energy captured by the primary ozonides. 
In principle, this could lead to a larger fraction of stabilized \ce{CH2=OO} than what is observed in the well studied ethylene ozonolysis reaction, which in turn could result in more formic acid. 
The C4 carbonyl oxides have similar pathways available to carbonyl oxide.
They may undergo unimolecular rearrangement and release OH radicals with continued fragmentation \cite{kuwata2005quantum,kuwata2008quantum,zhang2002mechanism,gutbrod1997kinetic}, or they could be collisionally stabilized. There were very few unidentified lines in our microwave spectra in the 12-26 GHz range, however, so likely concentrations of these species were quite low, and / or they fragmented to lighter species that fall outside the frequency range of the spectrometer (e.g. OH, CO).
\begin{scheme}[t]
\centering
\includegraphics[width=.45\textwidth]{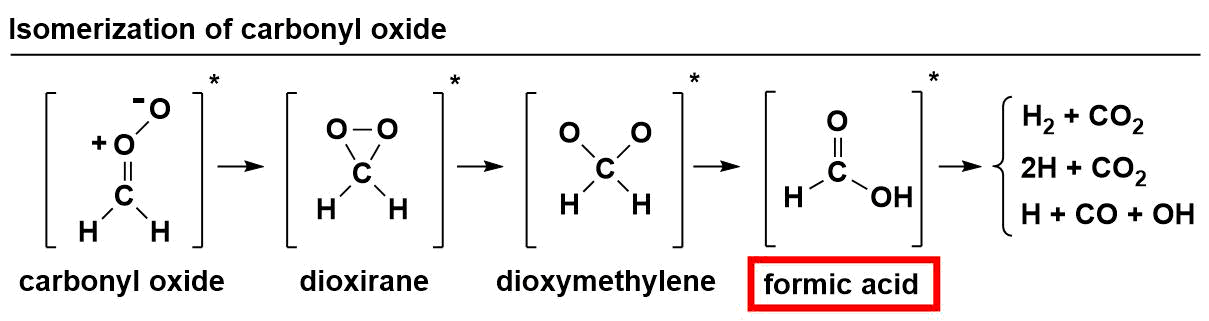}
\caption{Isomerization of the simplest Criegee intermediate, carbonyl oxide, to formic acid.}
\label{formicacid}
\end{scheme}

\section{Experimental}
\begin{figure}[b!]
\centering
\caption{Depiction of the cryogenic buffer gas cell coupled to the output of a nichrome heated alumina flow tube.}\label{4Kcell}
\includegraphics[width=0.5\textwidth]{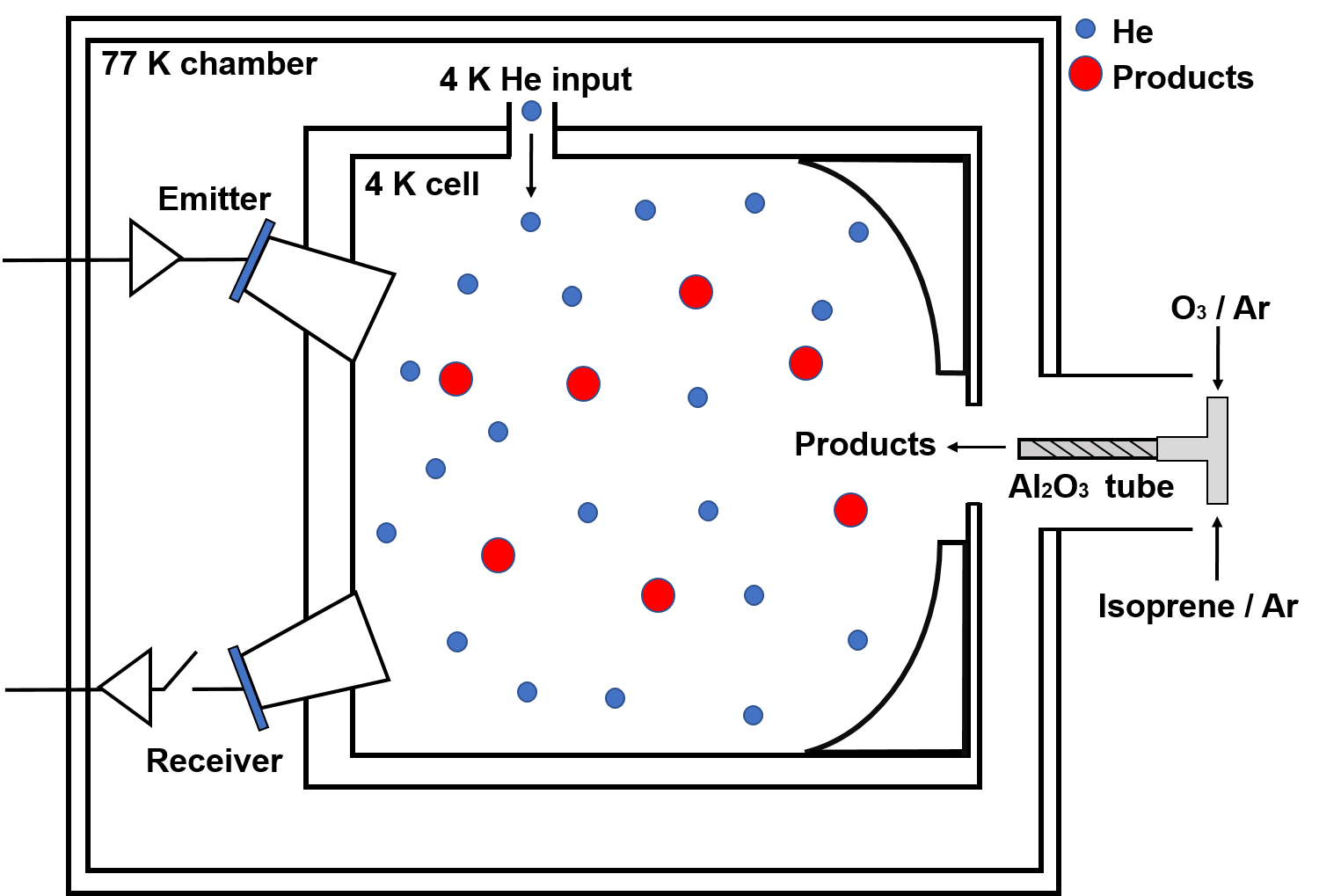}
\end{figure}

The buffer gas cell used here is similar to the one that has been described by Patterson and Doyle \cite{patterson2012cooling}. 
The active region of the buffer gas cell is a 19 cm x 19 cm x 19 cm vessel anchored to a 2 stage He cryostat, which cools the cell down to approximately 4-7 K (see Figure \ref{4Kcell}). 
This cell is surrounded by a second chamber held at 77 K by the first stage of the cryostat, with silicon diode temperature monitors in four locations throughout the cell (Lakeshore DT-670A1-SD). 
The helium buffer gas is pre-cooled and introduced into the cell at a flow rate of 7 sccm, creating an approximate density of 2 $\times 10^{14}$ He/cm$^{3}$.
Reagents are introduced behind the entrance to an alumina flow tube and allowed to mix for roughly 10 seconds before injection into the cryogenic buffer gas bath (cell aperture 19.05 mm).
As molecules enter the cell, they are slowed and cooled by collisions with He, which slightly warms the cell.
Molecules are considered fully thermalized after roughly 20-100 collisions\cite{patterson2012cooling}, with molecule - He collisions occurring roughly once every 10 $\mu$s (i.e. thermalization within less than 1 ms).
Observation times are 15-20 ms, limited by diffusion of the molecules to the cell walls.
Microwaves are injected into the active region of the cell at a repetition rate of 50,000 Hz.
A cryogenic switch is used to protect the receiver and cryogenic low noise amplifiers during microwave input. 

Dilute isoprene and ozone are introduced independently to the alumina flow tube (1/8" outer diameter, a 0.058" inner diameter, 9" length) using mass flow controllers and minimal 1/8" PFA tubing. 
The tube was heated with a nichrome wire wrap monitored by type K thermocouple to avoid build up of ice within the tube, and was maintained at 298 K for the branching ratios analysis.
The pressure (consistently about 5 Torr) was measured behind the entrance to the reactor by a Granville-Phillips 275 gauge.
The spectra presented below were collected in 60 MHz steps in triplicate to remove unwanted sideband interference from the mixing of the local oscillator and chirp, and utilized a 4.5 $\mu$s and 100 MHz chirp. 
Collecting a complete, continuous 12-26 GHz spectrum required approximately 5 hours, with 250,000-417,000 averages per 60 MHz step.
To improve signal to noise for branching ratio analysis, longer integrations were performed on individual peaks  (1,000,000-1,250,000 averages) using a more powerful, only 25 MHz wide chirp (still within the weak pulse limit). Branching ratio calculations as well as estimations for cryogenic buffer gas density, collision frequency, and residence time in the alumina tube are discussed in the Supplemental Information.
Line strengths necessary for branching ratio estimates were computed using PGOPHER \cite{western2017pgopher}.
Best fits for rotational parameters based upon experimentally observed lines for $^{13}$C-$trans$-isoprene and $^{13}$C-$ap$-MACR were calculated using the program Pickett \cite{pickett1998submillimeter}.

\section{Results and Discussion}

\begin{figure*}[t!]
\centering
\includegraphics[width=\textwidth]{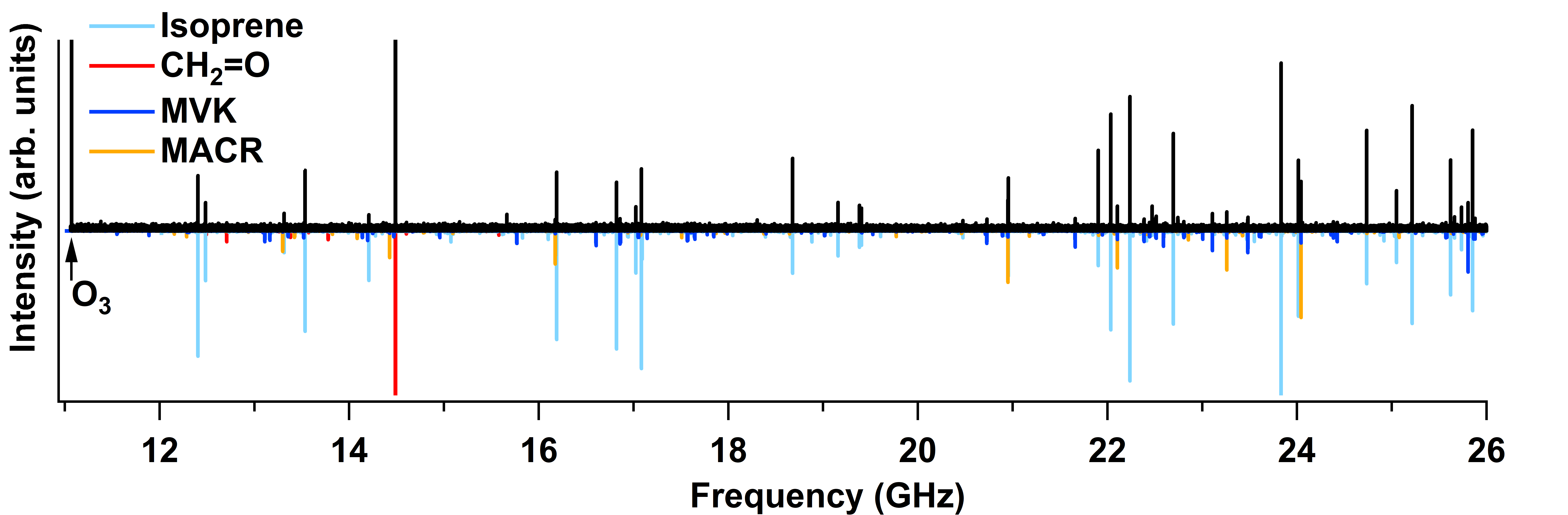}
\caption{The 12-26 GHz CP-FTMW spectrum of the ozonolysis of isoprene. The reactant ratio is 2:1 \ce{O3} to isoprene, and the total reagent concentration is 2\% in argon. 
}
\label{full}
\end{figure*}

Figure \ref{full} shows the full 12-26 GHz CP-FTMW spectrum that results when a 2:1 ozone to isoprene ratio in argon (2\% reagent total) is introduced behind the entrance to the heated alumina flow tube.
Authentic spectra of isoprene, formaldehyde, MACR, and MVK are displayed along the negative Y axis for reference. 
The most intense peak arises from formaldehyde, \ce{CH2=O} at 14488.48 MHz \cite{bocquet1996ground}, along with \ce{O3} at 11072.45 MHz \cite{gora1959rotational} and a number of isoprene peaks \cite{Lide1964microwave}.
The experimental conditions are chosen such that primary chemistry is dominant, as opposed to secondary chemistry such as oxidation of MVK or MACR.

\begin{figure}
\centering
\includegraphics[width=0.45\textwidth]{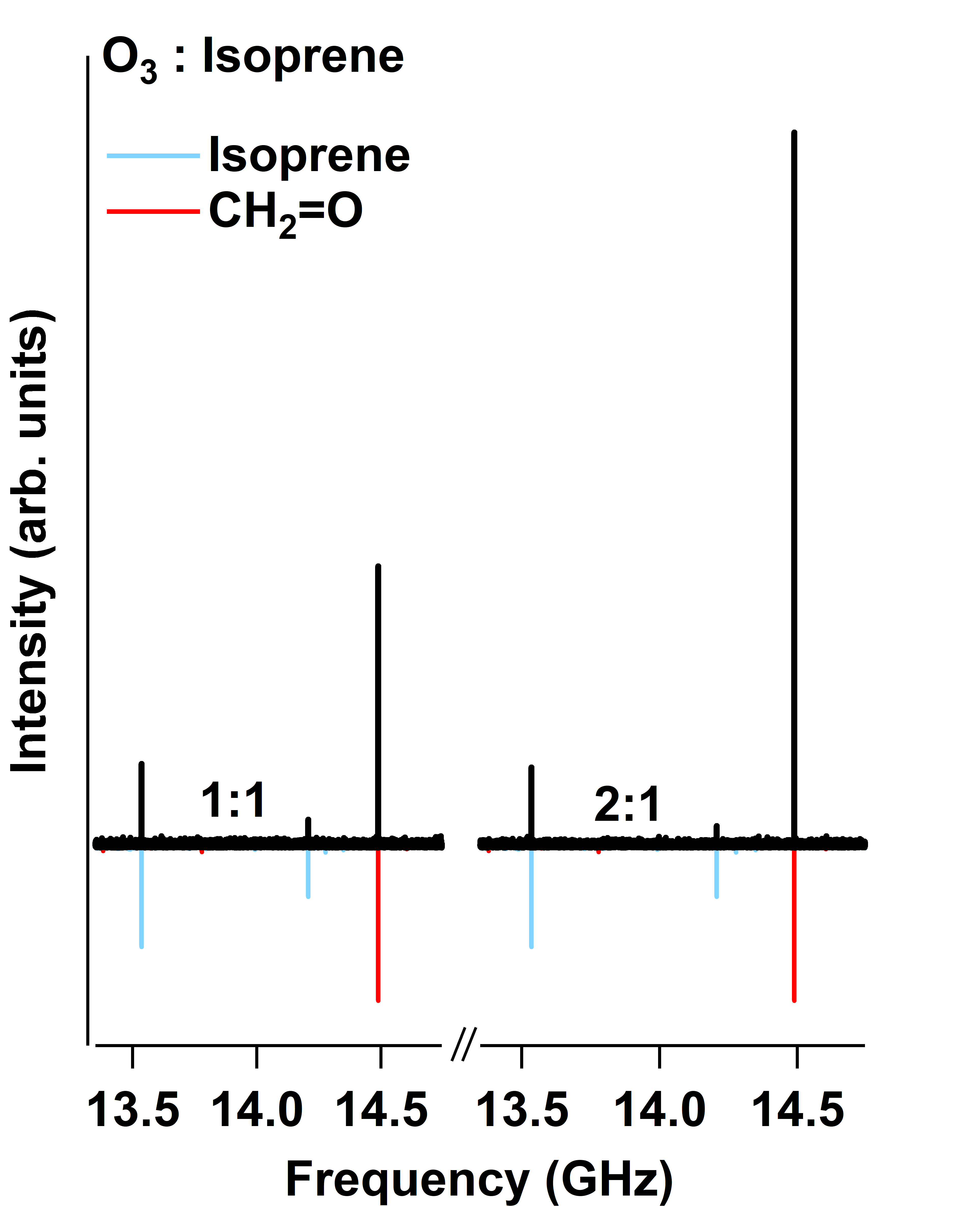}
\caption{Isoprene and formaldehyde (14.48 GHz) as the ratio of ozone to isoprene changed from 1:1 to 2:1. Spectra are scaled such that the isoprene peaks are identical in intensity. The large enhancement of the formaldehyde line indicates further extent of reaction under 2:1 conditions. 
}\label{2to1}
\end{figure}

\begin{figure}
\centering
\includegraphics[width=0.43\textwidth]{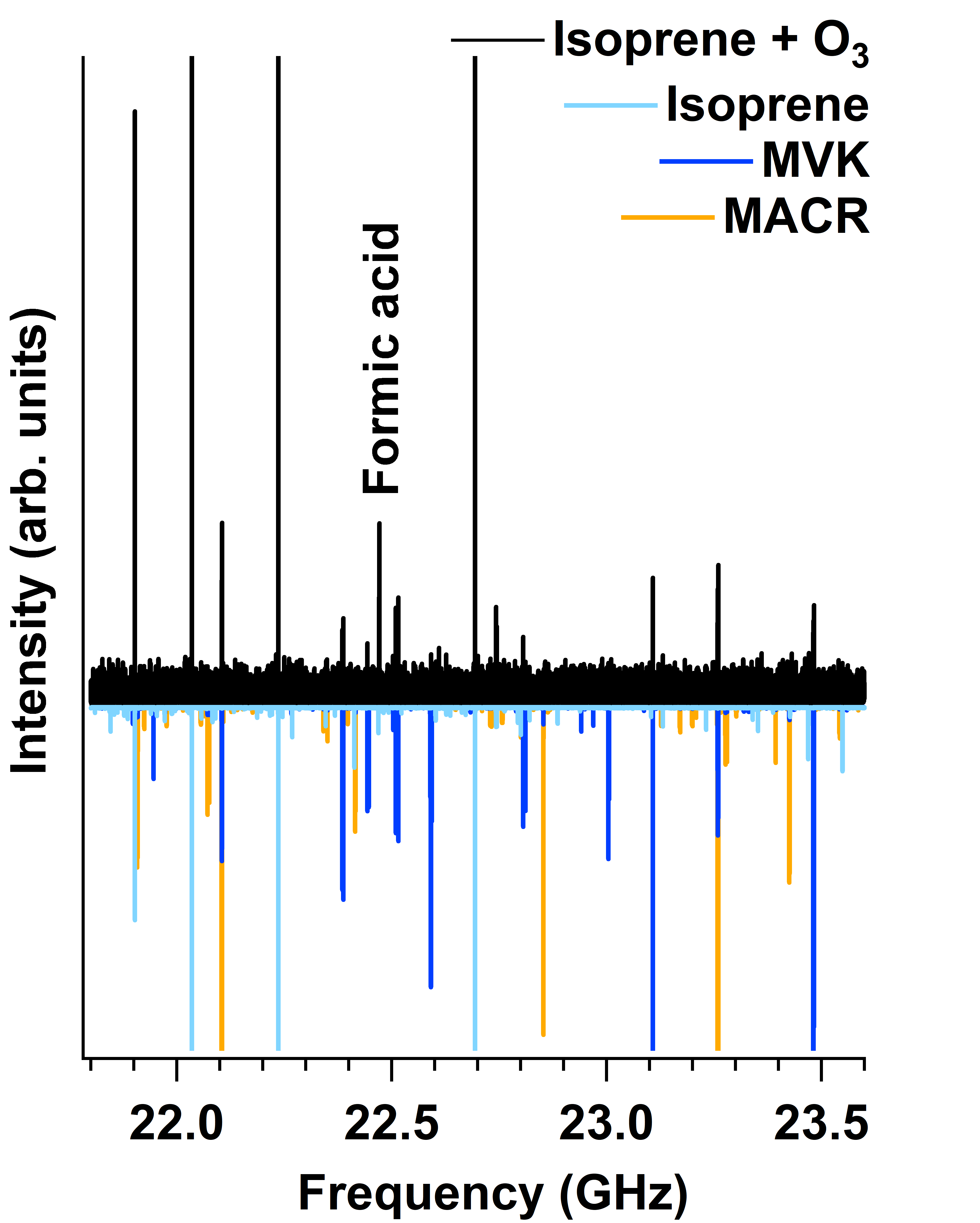}
\caption{Evidence for methacrolein (MACR), methyl vinyl ketone (MVK), and formic acid in the ozonolysis of isoprene. 
}\label{MACR}
\end{figure}

Figure \ref{2to1} displays how the extent of the reaction is controlled by simply varying the relative ratios of the two reactants; the left hand panel displays a 1:1 ozone to isoprene, and the right hand panel 2:1. 
The two spectra are scaled such that the isoprene peaks displayed at 13535.58 MHz ($5_{23} \rightarrow 5_{14}$) and 14207.15 MHz ($4_{13} \rightarrow 4_{04}$) are roughly equivalent in intensity.
Figure \ref{MACR} provides an expanded view of the features of Figure \ref{full} in the 22 - 23.5 GHz region.
Aside from isoprene itself, there is clear evidence for the oxidation products MACR and MVK in the 2:1 spectrum. 
Presumably, the presence of formic acid at 22471.18 MHz is a consequence of rearrangement of the Criegee intermediate \ce{CH2=OO} produced in tandem with both MVK and MACR, see Scheme \ref{formicacid}. 

The abundances of stable oxidation products in the ozonolysis of isoprene have been measured at two different concentrations of \ce{O3} and isoprene.
The ratios of MACR to MVK are summarized in Table 1 and compared to previous experimental results.
Although there was evidence for both $ap$- and $sp$-MVK, we found no evidence of $sp$-MACR or $cis$-formic acid (see Scheme \ref{conformers}).
Formaldehyde and formic acid each had only one line in the 12-26 GHz region, as opposed to MVK and MACR, where many lines (6-12) could be compared. 
Thus, the quantities of formic acid and formaldehyde are discussed in relative terms under different experimental conditions. 

\begin{table}[h!]
\centering\label{ratios}
\caption{The derived ratios of methacrolein to methyl vinyl ketone in the ozonolysis of isoprene compared to previous experimental values. For the ratios presented here, $ap$ and $sp$ MVK abundances were added for a total MVK and therefore MACR to MVK ratio. The reaction took place at 298 K and 5 Torr with 10 s residence time in the alumina flow tube prior to cooling and observation. Overall reagent concentrations of ozone to isoprene are 1:1 (4\% in Ar), and 2:1 (2\% in Ar).}
\begin{tabular}{lr}
\multicolumn{2}{c}{\textbf{Ratio of MACR to MVK}}\\
\hline
2.1 $\pm$ 0.4&	1:1 \ce{O3}:Iso this work\\
2.1 $\pm$ 0.2&	2:1 \ce{O3}:Iso this work\\
2.29& Kamens 1982 \cite{kamens1982ozone}\\
2.57 $\pm$ 0.02&	Grosjean 1993 \cite{grosjean1993atmospheric}\\
2.43 $\pm$ 0.03&	Aschmann 1994 \cite{aschmann1994formation}\\
2.44 $\pm$ 0.04&	Rickard 1999 \cite{rickard1999oh}\\
2.34 $\pm$ 0.03&	Nguyen 2016 \cite{nguyen2016atmospheric}\\
2.9 $\pm$ 0.2&		Ren 2017 \cite{ren2017investigation}\\
\hline
\end{tabular}
\end{table}
 
Utilizing the single strong line at 14488.48 MHz, we derived a formaldehyde/(MVK+MACR) abundance of 92\% and 88\% under 1:1 ozone to isoprene conditions and 2:1 ozone to isoprene conditions, respectively. 
These values are in relatively good agreement with previous reports\cite{grosjean1993atmospheric,aschmann1994formation} of 80 - 90\% . 
Because MVK and MACR are produced in tandem with \ce{CH2=OO}, which has the potential to isomerize to formic acid (see Schemes \ref{isoprene} and \ref{formicacid}), we can analyze the formic acid/(MVK + MACR) as a formic acid/(\ce{CH2=OO}) ratio.
By performing a similar analysis to formaldehyde using the single formic acid line at 22471.18 MHz, we derive an abundance of formic acid/(MVK+MACR) of 82\% and 70\% under 1:1 and 2:1 conditions, respectively.

At room temperature, 98\% of isoprene participating in ozonolysis is expected to be in the $trans$ form, which would lead to $ap$-MVK, see Scheme \ref{conformers}.
However, as the molecules collide with the walls of the reactor and argon carrier gas on their way to the cold cell, they appear to redistribute back towards their room temperature populations.

Using the W1BD thermochemical method\cite{barnes2009unrestricted}, we calculated the 298 K enthalpies and free energies of $trans$- and $gauche$-isoprene, $sp$- and $ap$-MVK, and $sp$- and $ap$-MACR .
By the W1BD method, we obtained $\Delta_{\textrm{rxn}}$G$_{298}$ = 1.42 kJ/mol and $\Delta_{\textrm{rxn}}$H$_{298}$ = 1.84 kJ/mol for [$ap$-MVK\ce{<=>}$sp$-MVK], corresponding to an anticipated $sp$-MVK abundance of 36\% at room temperature ($\Delta$G = $\Delta$H - T$\Delta$S = -RT ln(K$_{\textrm{eq}}$); K$_{\textrm{eq}}$ = [$sp$-MVK]/[$ap$-MVK]).  
This is a reasonable estimate given the only previously reported experimental value to our knowledge of $\Delta_{\textrm{rxn}}$H$_{298}$ = 2.34 $\pm$ 0.2 kJ/mol \cite{bowles1969conformations}, which would correspond to roughly 31\% $sp$-MVK abundance.
There was no evidence for $sp$-MACR, which is consistent with the W1BD result that ($ap$ \ce{<=>} $sp$) $\Delta_{\textrm{rxn}}$H$_{298}$= 13.7 kJ/mol and $\Delta_{\textrm{rxn}}$G$_{298}$ = 13.0 kJ/mol, resulting in a room temperature population of about 0.5\% $sp$-MACR.
At the W1BD level of theory, the $gauche$ form of isoprene is the higher energy isomer of the two ($\Delta_{\textrm{rxn}}$H$_{298}$ = 11.3 kJ/mol; $\Delta_{\textrm{rxn}}$G$_{298}$ =  10.5 kcal/mol). 
We estimate that $gauche$-isoprene exists at 1.6\% abundance at room temperature, and hence should be present in trace quantities.
Previously unreported $^{13}$C lines and spectroscopic parameters for $ap$-MACR and $trans$-isoprene are summarized in the Supplemental Information, and additional work is underway to characterize $gauche$-isoprene from these buffer gas experiments.

Figures \ref{full}-\ref{MACR} provide a visual example of the level of sensitivity and control that can be achieved when the reactor is coupled to a cryogenic buffer gas cell.
Although the extent of reaction was not large, reliable quantitative data for relative abundances were extracted from very short reaction times and at low concentrations.
In order to determine extent of reaction, $gauche$-isoprene must be characterized spectroscopically, a task which has proven challenging and which will be the topic of a future report.
The ability to exert control over temperature, pressure, concentration, and therefore extent of reaction under these conditions suggests this approach might be useful for kinetic analysis.
Our present analysis is limited for purely technical reasons- the finite bandwidth of the microwave instrument. 
Since formic acid and formaldehyde each only have one rotational line in the region of the spectrometer, it is difficult to reliably assess the experimental error of their relative abundances. 

Though a number of studies have analyzed product distribution in the ozonolysis of isoprene, the extent to which ozone favors addition at either site in isoprene remains a topic of debate due to the presence of formaldehyde in multiple channels\cite{aschmann1994formation,grosjean1993atmospheric,rickard1999oh,nguyen2016atmospheric,ren2017investigation,kamens1982ozone}.
The detection methods of choice for previous studies have been various forms of gas chromatography \cite{aschmann1994formation,rickard1999oh,nguyen2016atmospheric,ren2017investigation} and Fourier-transform infrared spectroscopy \cite{nguyen2016atmospheric,ren2017investigation}, with laser induced fluorescence \cite{nguyen2016atmospheric} and high performance liquid chromatography \cite{grosjean1993atmospheric} being used less frequently.
Lengthy residence times on the order of hours inside Teflon reaction chambers were required to build up detectable quantities of products.
Due to this constraint, extensive kinetic analysis was needed to compensate for additional chemistry such as \ce{O3 +} MVK or MACR.
Previous experimental results present values between 2.29 - 2.9 for the ratio of MACR to MVK, a range that brackets our results of 2.1 $\pm$ 0.4 and 2.1 $\pm$ 0.2 within the margins of error (see Table 1). 

Given the formation of radicals in this reaction and their rapid reaction rates, the influence of side reactions on the reported abundances must be considered.
The most prominent collision partners aside from argon present in the reactor were isoprene and ozone. 
The respective reaction rates of ozone with isoprene, MVK, and MACR according to Atkinson (12.3, 4.41, and 1.08 x 10$^{-18}$ cm$^3$ molecule$^{-1}$s$^{-1}$, respectively) are six orders of magnitude slower than the reaction rate of \ce{CH2=OO} with ozone (1 x 10$^{-12}$ cm$^3$ molecule$^{-1}$s$^{-1}$) \cite{atkinson1994gas, vereecken2014reactions}.
It is possible that there is a minor contribution of formaldehyde from the rapid \ce{CH2=OO} + \ce{O3} $\rightarrow$ \ce{CH2=O} + \ce{2O2}, given the slight increase in observed formaldehyde under more concentrated conditions (92\% at 4\% reagent, 88 at 2\% reagent). 
However, we cannot say this definitively as \ce{O2} cannot be observed
by microwave spectroscopy as an indicative side product.
We must also consider the influence of OH radical chemistry on the ratios reported, as it has been well documented that alkene ozonolysis is one of the most important non-photochemical sources of tropospheric OH radical \cite{gutbrod1997kinetic,zhang2002mechanism}.
Isoprene and OH radical react very rapidly (k roughly 1 x 10$^{-10}$ molecule$^{-1}$s$^{-1}$), generating large fractions of MVK (0.27), MACR (0.19), and formaldehyde (0.50), with a small branching ratio to 3-methylfuran (0.038) \cite{mcgivern2000experimental}.
This reaction would then yield a ratio of  MACR to MVK of 0.7, and would therefore skew the derived ratio of [MACR]/[MVK] to a lower value.

The highest ratio of MACR to MVK was observed recently by Ren and coworkers who studied the ozonolysis of isoprene in a large, outdoor simulation chamber \cite{ren2017investigation}. In the presence of an OH radical scavenger cyclohexane, they observed a yield of 42 $\pm$ 6\% MACR and 18 $\pm$ 6\% MVK, or a ratio of 2.9 $\pm$ 0.2 MACR to MVK. 
Ren $et.$ $al.$ reported a rate constant of MVK with ozone comparable to Atkinson, however they reported a rate constant of MACR with ozone (7.1 $\pm$ 0.6 x 10$^{-19}$ cm$^3$ molecule$^{-1}$ s$^{-1}$) which is significantly lower than Atkinson. 
If correct, this would result in under-estimation of MACR consumed in secondary chemistry, and in turn would lead to the resulting higher ratio reported.
We searched for anticipated ozonolysis products of MVK and MACR and saw no evidence for these species (namely methylglyoxal and hydroxylmethyl hydroperoxide)\cite{chen2008aqueous}. 
As such, this relatively slow (k about 1 x 10$^{-18}$ cm$^3$ molecule$^{-1}$s$^{-1}$) secondary chemistry does not appear to play a significant role in affecting the MACR to MVK ratio under our experimental conditions.

It was difficult to find comparable data for production of formic acid from the ozonolysis of isoprene, as previous studies that do mention formic acid discuss its production as a function of humidity for atmospheric implications \cite{neeb1997formation, horie1994formation, paulot2009isoprene}.
We are confident that the formic acid observed in this experiment was a result of \ce{CH2=OO} isomerization and not the bimolecular reaction of \ce{CH2=OO + H2O}. 
In reactions with water, hydroxymethyl hydroperoxide (HMHP, \ce{HOCH2OOH}) is the most abundant product, followed by formic acid and hydrogen peroxide.
We found no evidence in our spectra\cite{nguyen2016atmospheric} for either HMHP or \ce{H2O2}.
The dipole moment\cite{kim1962dipole} along the principle axis of formic acid is 1.391 $\pm$ 0.005 D, and that of HMHP is predicted by Nakajima $et$. $al.$ to be 1.89 D, indicating that our instrument has adequate sensitivity to detect the presence of either species \cite{nakajima2015observation}.
Although we do not have multiple experimental lines to report for our formic acid ratios, it appears that a significant fraction of \ce{CH2=OO} isomerizes to HCOOH under the given conditions (70-80\%).
Large products MVK and MACR are in principle capable of removing large quantities of excess energy from the reaction that would otherwise be imparted to \ce{CH2=OO}.
This stabilization could lead to a much larger quantity of formic acid production than anticipated in the ozonolysis of smaller, terminal alkenes.

We believe that the major contribution of experimental error (roughly 20\%) is due to instrument response variations across the 12-26 GHz microwave circuit. 
Lines were tested for internal consistency prior to use in ratios analysis, meaning multiple lines for one species (when possible) first had to pass a 1 to 1 ratios test, within $\pm$ 20\%.
There is detailed information on this analysis method in the Supplemental Information.
This could potentially be addressed with a few different approaches. 
First, there are three apparent regions in the 12-26 GHz spectrum with distinct noise levels. 
One in principle could record a blank spectrum without molecule signal to normalize the noise level, however this would be necessary under each new set of experimental conditions (i.e. number of averages, chirp conditions etc. must be identical).
Second, one could also check the power transmission by putting in a known amount of power, but this must be kept very low in order to protect the receiver circuit low noise amplifiers. 
Third, one could normalize utilizing a well characterized molecule with many transitions throughout the instruments bandwidth, however this has clear limitations.
Lastly, it is likely that there is minor error contribution from temperature drift throughout the time of data acquisition (about 0.1 K per hour). This could be addressed with individual temperature measurements for each intensity measured for analysis.

With the identification of all stable initial oxidative products in the ozonolysis of isoprene in the 12-26 GHz region, we now intend to extend this work towards observation of the Criegee intermediates predicted in Scheme \ref{isoprene}. 
The given setup would need modification for radical observation, as evidenced by lack of unidentified lines between 12-26 GHz (therefore no unidentified Criegee radicals).
This may be achieved by decreasing the diameter of the alumina tube to increase the reaction pressure; this in turn would allow for more collisional stabilization of the Criegee intermediates prior to entering the buffer gas cell. It is also likely that total reagent concentrations slightly higher than 4\% in argon could be used, enhancing overall signal to noise of trace species.
A straightforward modification could also involve a pick-off source, in which a small portion of the reaction is fed from the flow tube into the buffer gas cell with the remainder being pumped away, allowing for much higher overall flow rates and pressures. 
Observation of Criegee intermediates could also be achieved with the approach of Womack $et.$ $al.$ in which a modified pulsed valve was used to observe \ce{CH2=OO} in the reaction of ozone and ethylene \cite{womack2015observation}.
In that experiment, reagents were introduced via separate capillaries fed directly behind the exit of the pulsed valve, with the exhaust at atmospheric pressure and high flow rates for constant reagent replenishment.
Alternatively, \ce{CH2=OO} has been observed via discharge experiments in cavity microwave experiments, and it would be straight forward to couple the discharge to the buffer gas cell in the hunt for new species \cite{mccarthy2013simplest}.
An additional, straightforward extension of this work would be to do a complete branching ratios analysis, which to our knowledge has never been achieved experimentally. However in the ozonolysis of isoprene, formaldehyde is produced indistinguishably by two pathways. With simple $^{13}$C isotopic substitution at the 1 or 4 position in isoprene, these pathways could be distinguished by comparison of \ce{CH2=O} and $^{13}$\ce{CH2=O} using the methods described here.

\section{Conclusion}

We have developed a new technique to determine reaction product ratios utilizing high resolution microwave spectroscopy; this lays the foundation for straightforward, complete branching ratios analysis with simple isotopic substitution. We report previously uncharacterized rotational constants for $^{13}$C-$ap$-MACR and $^{13}$C-$trans$-isoprene, and have derived ratios of MACR to MVK of 2.1 $\pm$ 0.4 under 1:1 ozone to isoprene (4\% in argon) and 2.1 $\pm$ 0.2 under 2:1 ozone to isoprene (2\% in argon), in fair agreement with the literature.
In this experiment the speed of reagent mixing time (about 10 s), low concentrations utilized ($<$ 4\%), high resolution, and the high sensitivity that can be achieved with cryogenic electronics (ppb scale) have combined to give comparable results reported from previous experimental studies by far simpler analytical means.
For these reasons we believe this instrument is well suited for studying complicated mixtures and a wide variety of chemical reactions.

\section*{Conflicts of interest}
There are no conflicts to declare.

\section*{Acknowledgements}
This work was supported by NSF grant CHE-1566266, as well as the award for abstract 1555781 IDBR type A. 
S. E. acknowledges funding through the Schr\"{o}dinger Fellowship of the Austrian Science Fund (FWF): J3796-N36. We also thank Kirill Prozument and John F. Stanton for insight on product ratio calculations, and John W. Daily for helpful discussions on cell density and residence time estimations.

\balance

\clearpage
\providecommand*{\mcitethebibliography}{\thebibliography}
\csname @ifundefined\endcsname{endmcitethebibliography}
{\let\endmcitethebibliography\endthebibliography}{}

\end{document}